\documentclass[aps,prl,twocolumn]{revtex4-2}
\usepackage{amsmath, amssymb, graphicx, xcolor, float}

\graphicspath{{./figures/}} 

\begin{document}

\title{Thermalization slowing down in multidimensional Josephson junction networks}
\author{Gabriel M.~Lando}
\author{Sergej Flach}
\affiliation{Center for Theoretical Physics of Complex Systems, Institute for Basic Science (IBS), Daejeon, Korea, 34126}
\keywords{thermalization slowing-down, Lyapunov spectrum, Josephson junctions, ergodic hypothesis}

\begin{abstract}
We characterize thermalization slowing-down of Josephson junction networks in 1, 2 and 3 spatial dimensions for systems with hundreds of sites by computing their entire Lyapunov spectra. The ratio of Josephson coupling $E_J$ to energy density $h$ controls two different universality classes of thermalization slowing-down, namely the weak coupling regime, $E_J/h \rightarrow 0$, and the strong coupling regime, $E_J/h \rightarrow \infty$. We analyze the Lyapunov spectrum by measuring the largest Lyapunov exponent and by fitting the rescaled spectrum with a general ansatz. We then extract two scales: the Lyapunov time (inverse of the largest exponent) and the exponent for the decay of the rescaled spectrum. The two universality classes, which exist irrespective of network dimension, are characterized by different ways the extracted scales diverge. The universality class corresponding to the weak-coupling regime allows for the coexistence of chaos with a large number of near-conserved quantities and is shown to be characterized by universal critical exponents, in contrast with the strong-coupling regime. We expect our findings, which we explain using perturbation theory arguments, to be a general feature of diverse Hamiltonian systems.
\end{abstract}

\keywords{Thermalization, Josephson junctions, chaos and ergodicity, Lyapunov exponents, integrability}

\maketitle

Thermalization has been a central concept in theoretical physics since the birth of statistical mechanics. It allows for an explanation as to how time-dependent microscopic dynamics gives birth to macroscopic equilibrium, and is deeply rooted in the classical concepts of chaos and ergodicity \cite{gibbs1902elementary, boltzmann2022lectures, aleksandr1949mathematical, fermi1955studies, landau2013statistical, gallavotti1999statistical}. Here, the assumption that all arbitrary initial conditions lead to the same long-time average when dynamics is sufficiently ergodic -- and, therefore, that time averages and ensemble averages are interchangeable -- is fundamental. It is then natural to wonder what happens when systems that obey this \textit{ergodic hypothesis} approach a limit in which it ceases to be valid. In the context of many-body Hamiltonian dynamics, which lies beyond the applicability horizons of Kolmogorov-Arnol'd-Moser (KAM) theory \cite{arnol2013mathematical, de1988hamiltonian, poschel2000, wayne1984kam}, this matter can be envisaged as the study of how chaotic many-body systems approach integrability.

Since thermalization is reflected in the behavior of long-time averages, investigations so far have privileged its diagnose by computing classical expectation values \cite{mithun2019dynamical, mithun2021fragile, inarrea2022chaos}. This introduces unnecessary moving parts, \textit{e.g.}~choices of coordinates and classical observables, with ``good'' choices often obscured \cite{baldovin2021statistical}. As a consequence, recent work has instead focused on Lyapunov spectra (LS): Contrary to observables, LS are coordinate-independent and invariant under a wide range of transformations \cite{eichhorn2001transformation}, providing a complete measure of chaos and its characteristic time scales for any classical system \cite{constantoudis1997nonlinear, iubini2021chaos,malishava2022lyapunov}. In a recent paper they were employed to show that a specific class of discrete 1-dimensional unitary maps, similar to nonlinear quantum random talks, thermalize \textit{via} two different pathways depending on whether the target integrable limit decouples interactions in real or reciprocal space \cite{malishava2022lyapunov}. However, due to substantial computational challenges, little to nothing is currently known about how many-body Hamiltonian systems cease to thermalize, and whether or not their thermalization slowing-down processes fit the universality classes discovered in \cite{malishava2022lyapunov}.

In this Letter, we characterize the thermalization slowing-down of classical Hamiltonian systems with hundreds of degrees of freedom, addressing the issue mentioned above and showing these universality classes are a general property of Hamiltonian lattices in any dimension. Our analysis is based on a direct computation of the full LS, for which we introduce an ansatz capable of fitting the spectra in any perturbation regime. The models chosen are 1-, 2- and 3-dimensional Josephson junction networks, which have been extensively studied in 1d as a probe to the richness of 1-dimensional physics \cite{geerligs1989charging, chow1998length, pop2010measurement, mulansky2011strong, pino2016nonergodic, cedergren2017insulating}, and in 2d as an exact analogue to the XY-chain, offering the possibility of simulating a multitude of statistical phenomena in chips \cite{newrock2000two, martinoli2000two}. Our calculations show that the models thermalize according to two different universality classes near their integrable limits depending on the network type -- either short-ranged or long-ranged (essentially all-to-all), spanned by the weak nonintegrable perturbation over the set of actions (which are integrals of motion at the very limit \cite{mithun2021fragile,supmat}). In the long-range network (LRN) regime the LS is always well-approximated by an analytical function that depends only on the maximal Lyapunov exponent (mLE). The short-range network (SRN) regime is drastically different, with LEs decreasing exponentially according to a second diverging scale, the \textit{spectral decay rate}, rendering thermalization extremely slow when compared to the Lyapunov time (the inverse of the mLE).

Both the Lyapunov time and the spectral decay rate are shown to be remarkably universal in the SRN class through the extraction of scaling laws and their critical exponents. The LRN class is not characterized by additional universal scaling laws, a phenomenon we explain by invoking classical perturbation theory. In the SRN class, near-zero and large LEs coexist, such that regular submanifolds are gradually reconstructed within a sea of strong chaos as the integrable limit is approached. Such coexistence is absent in the LRN class, in which chaos vanishes almost evenly and tori are rebuilt nearly all at once. This suggests that the transition to integrability in the SRN and LRN classes, if interpreted as a phase transition, can be of more than one type (\emph{e.g.}~first or second order).

Our findings can be used to interpret equipartition, thermalization and ergodicity both in experimental and theoretical investigations, and we expect them to be useful in the analysis of any system displaying long- and/or short-range interactions, and many-body localization.

We shall now introduce the systems employed throughout the rest of this Letter. Denoting momentum and position by $\mathbf{p}$ and $\mathbf{q}$, the Hamiltonians for all networks considered here have the general form
\begin{equation}\label{eq:pre_hamiltonian}
    H(\mathbf{p},\mathbf{q}) = \frac{\vert\mathbf{p}\vert^2}{2} + E_J \sum_{\langle \sigma_1, \sigma_2 \rangle} \big[1 - \cos ( q_{\sigma_1} - q_{\sigma_2} ) \big] \, ,
\end{equation}
where $E_J$ is the Josephson coupling between two neighboring superconducting islands, and $\sigma_1, \, \sigma_2 $ are multi-indices counting nearest neighbors. Associated to each component of position there is a momentum $p$ entering the kinetic term, such that $p^2/2$ is interpreted as the island's Coulomb charging energy, and the rotor angle $q$ as the phase of the superconducting order parameter. For the 1d network we assume periodic boundary conditions, while for the 2 and 3d cases -- which we take as a square and a cube lattices with edge $N$ -- we leave them open. We shall refer to different dimensionalities implicitly by writing the total number of sites as $N$, $N^2$ and $N^3$ for $d=1,2$ and 3, \textit{e.g.~}the circular ring with 120 sites and the $5\times5\times5$ cube will be referred to as $N=120$ and $N=5^3$.

The Josephson junction networks \eqref{eq:pre_hamiltonian} conserve both total momentum and energy. The former can always be set to zero due to the Galilean invariance of macroscopic systems, or periodic boundary conditions. The relevant parameters are then the energy density, $h=H/N$ and $E_J$. Due to the boundedness of the potential energy term in \eqref{eq:pre_hamiltonian}, $h \to 0$ can only be achieved for fixed $E_J$ by constraining $q_{\sigma_1} - q_{\sigma_2}$ and $\mathbf{p}$ to be small. The cosine can then be expanded as $\cos x \approx 1 - x^2/2$ and, for small energies, the networks become integrable sets of coupled harmonic oscillators. The same effect can be realized as $E_J \to \infty$ while keeping $h$ fixed, such that $E_J/h \gg 1$ takes care of both possibilities and furnishes what we refer to as the LRN regime, in which sites remain strongly connected. The opposite case $E_J/h \ll 1$ produces the SRN regime, in which the sites become weakly coupled sets of rotors or nearly disconnected superconducting islands.

We enter the SRN and LRN regimes by varying different parameters, namely $(E_J \to 0, h=1)$ for the first and $(E_J=1,h \to 0)$ for the latter, with the data continuously glued together using the simple rescaling transformations described in \cite{supmat}. This is only possible because the Hamiltonian \eqref{eq:pre_hamiltonian} has effectively a single parameter, namely $E_J/h$, in terms of which we will present our results \cite{sadia2022prethermalization}. 

\begin{figure}
    \centering
    \includegraphics[width=\linewidth]{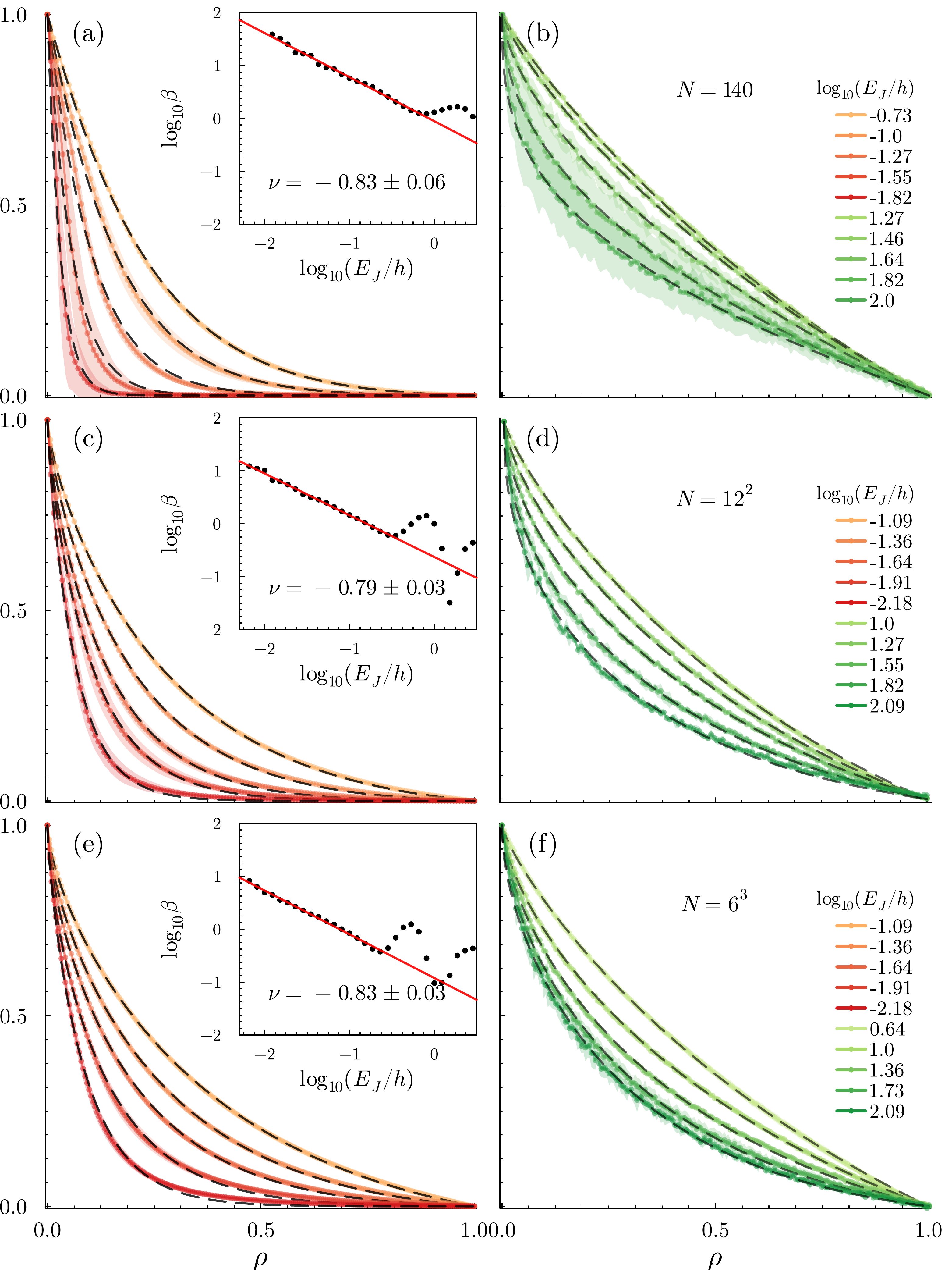}
        \caption{Selected sets of rescaled Lyapunov spectra $\bar{\Lambda}(\rho)$ in the short-range regime (red and orange tones) for the (a) 1d, (c) 2d and (e) 3d networks, together with the fits obtained from the ansatz \eqref{eq:fit_ansatz} (black dashed curves). The variable $\rho$ is a normalized counting index. The insets show the $\beta$ coefficients in \eqref{eq:fit_ansatz}, which measure the speed of exponential decay, together with a power-law fit with exponent $\nu$. (b, d, f) Same, but in the long-range regime in which no exponential decay is visible and the Lyapunov exponents do not bend towards zero. For all plots, error bars are computed from runs with different initial conditions and shown as ribbons \cite{supmat}, and the values of $E_J/h$ and system sizes/dimensionalities are shown in the panels on the right.}
        \label{fig:spectra}
\end{figure} 

\begin{figure}
    \centering
    \includegraphics[width=\linewidth]{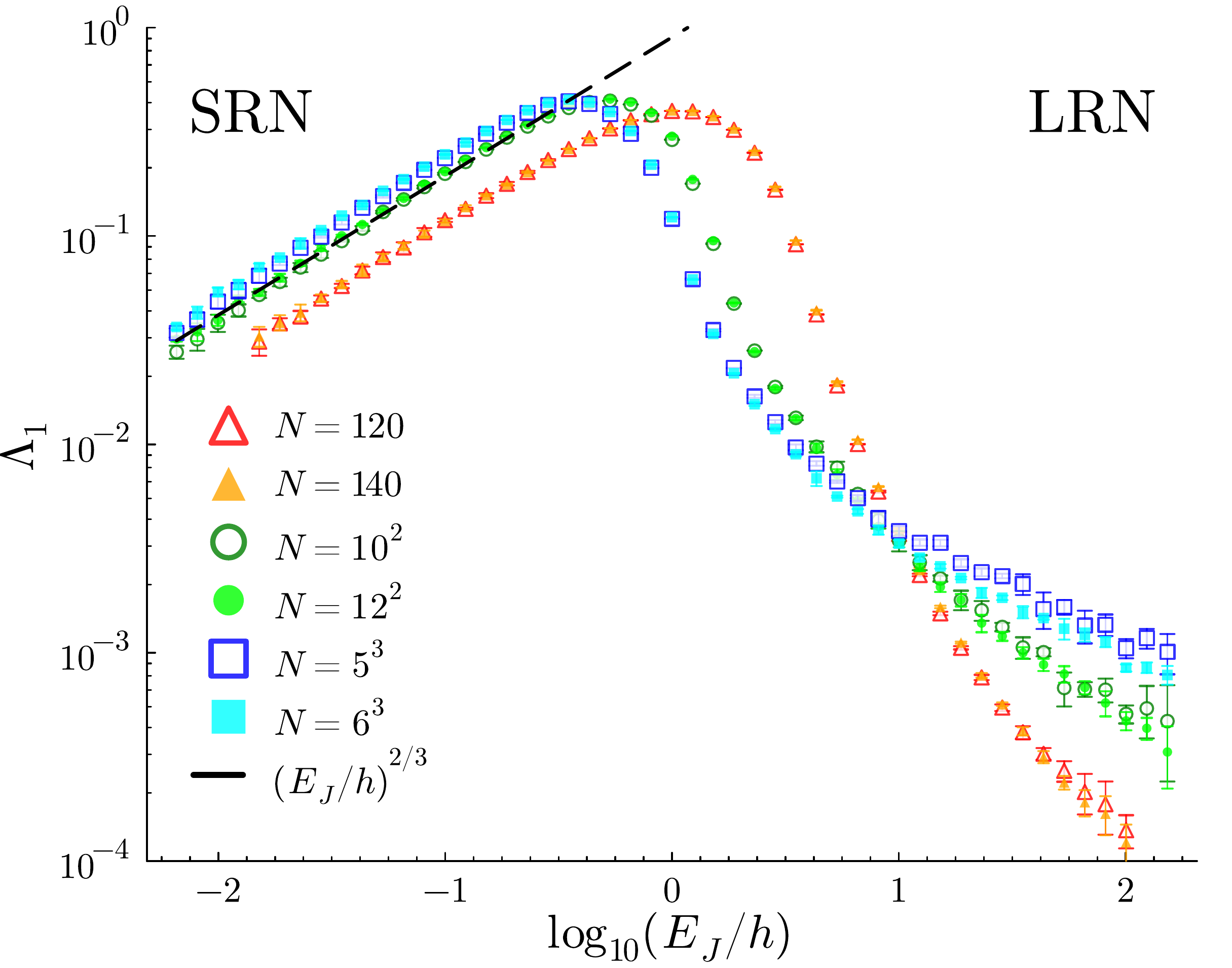}
        \caption{Maximal Lyapunov exponent versus $\log E_J/h$ for all networks. Solid markers denote the same configuration as their open counterparts, but with a larger number of sites. The solid black line shows a power-law with critical exponent $2/3$, predicted analytically in \cite{casetti1996riemannian}.}
        \label{fig:lambda1s}
\end{figure}

Since we are interested in the vicinity of integrable limits, the dynamics is weakly chaotic and the time necessary to resolve LEs is very long. This reliance on long times, together with the fact that we would like to preserve conserved quantities as much as possible in order to avoid creating ``fake'' chaos due to numerical instabilities, lead to the need of employing symplectic integrators \cite{yoshida1990construction, mclachlan1992accuracy, benettin1994hamiltonian, hairer1997life}. Here, we employ the 2nd order optimized method of \cite{mclachlan1992accuracy} for solving the equations of motion for the monodromy matrix $\mathcal{M}$, \textit{i.e.}~the Jacobian of the Hamiltonian flow, $\Phi$:
\begin{equation}\label{eq:tangent_eqs}
    \frac{\text{d}\mathcal{M}(\mathbf{z};t)}{\text{d}t} = \mathcal{J} \text{Hess} \, H(\mathbf{z}') \bigg|_{\mathbf{z}' = \Phi (\mathbf{z};t)} \mathcal{M}(\mathbf{z};t) \, .
\end{equation}
In the above, $\mathbf{z}=(\mathbf{p},\mathbf{q})$ denotes a $2N$-dimensional phase-space variable composed of $N$ momentum-position pairs, $H$ is the system's Hamiltonian, and $\mathcal{J}$ is the symplectic matrix \cite{arnol2013mathematical, de1988hamiltonian}. The initial point $\mathbf{z}$ is evolved to $\mathbf{z}' = \Phi(\mathbf{z};t)$ by the Hamiltonian flow, which is computed in parallel. We then extract the LS from $\mathcal{M}$ by employing the well-known QR-decomposition algorithm of \cite{benettin1980lyapunov}. This method, which is both accurate and efficient for continuous systems \cite{geist1990comparison, mogavero2023timescales}, works especially well if small times steps, $\mathrm{d}t$, are used. This is what lead us to choose a low-order algorithm with small $\mathrm{d}t$ instead of the standard prescription, \emph{i.e.}~a high-order one with large $\mathrm{d}t$ \cite{wanner1996solving, laskar2001high, mogavero2023timescales}. Here, we fix the step size as $\text{d}t = 0.2$ for all models considered, with final propagation times of $t_f = 2 \times 10^6$ for 1d and 2d systems, and $10^6$ for the 3d case. This results in a relative energy and total momentum drifts no larger than $10^{-3}$ and $10^{-12}$, respectively, with small variations depending on the number of sites and dimensionality \cite{supmat}.

Since the computed LS will be very different depending on the system's parameters, we rescale them in order to facilitate comparisons. The rescaling transformation for each LE is simply $\overline{\Lambda}_i = \Lambda_i/\Lambda_1$, where $\Lambda_1 > \Lambda_2 > \dots > \Lambda_{N-1} = \Lambda_N=0$ are the original LEs, the last two being equal to zero due to total momentum and energy conservation. Since we work with systems with several different sizes, we also define a rescaled $x$-axis, namely $\rho = \{ i/N\}_i$. This results in plots for the rescaled spectra that vary between 0 and 1 both in the $y$ as in the $x$ axes, as can be seen in the set of LS displayed in Fig.~\ref{fig:spectra}. The $E_J/h$ values and the number of sites for each row can be seen as insets in the panels on the right. We also keep track of the mLEs, which are shown for several system sizes and dimensionalities in Fig.~\ref{fig:lambda1s}. 

\begin{figure*}
    \centering
    \includegraphics[width=0.8\linewidth]{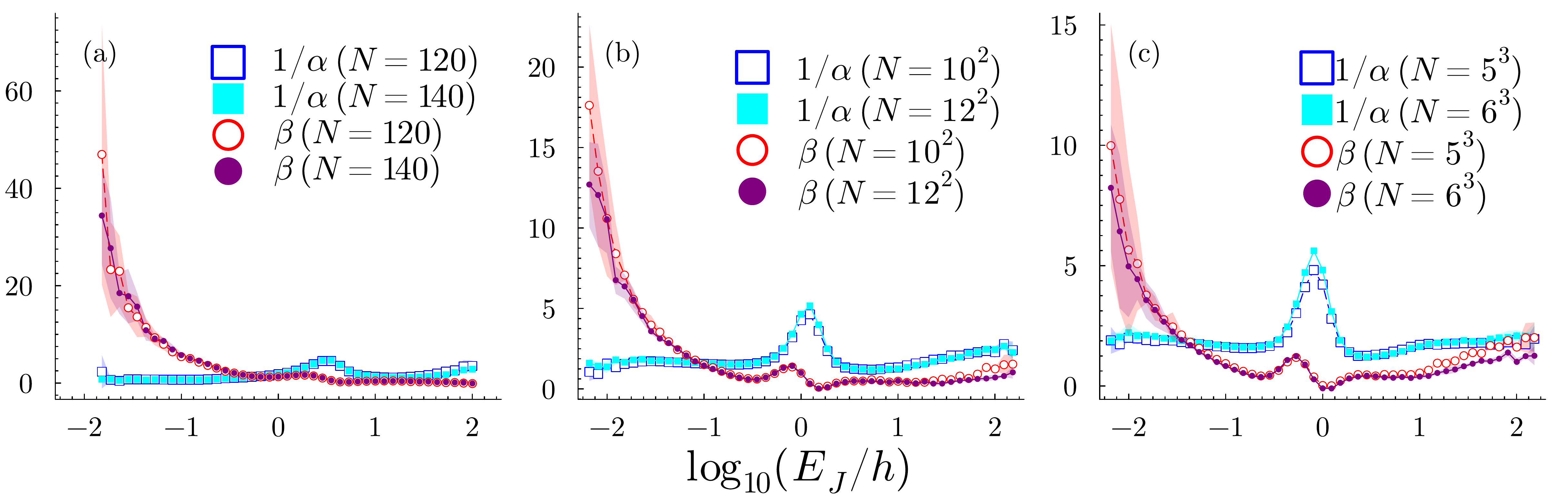}
        \caption{Fit coefficients for the ansatz \eqref{eq:fit_ansatz} for two sizes of the (a) 1d, (b) 2d and (c) 3d networks. A log-log plot which clearly displays the power-law behavior of the $\beta$-coefficients can be seen in the insets of Fig.~\ref{fig:spectra}. Note that we plot $1/\alpha$ instead of $\alpha$ for ease of visualization. Error bars are displayed as ribbons.}
        \label{fig:as_and_bs}
\end{figure*}

In order to provide a more quantitative distinction between SRN and LRN regimes, we employ a modification of an ansatz suggested in \cite{iubini2021chaos} in which the rescaled LS away from integrable limits is approximated as $1-\rho^\alpha$. Our modification arises from noticing that LS in the SRN regime appear to decay exponentially as a function of $\rho$, such that the new ansatz 
\begin{equation}\label{eq:fit_ansatz}
    \overline{\Lambda(\rho;\alpha,\beta)} = (1-\rho^\alpha) \, e^{-\beta \rho} \quad 
\end{equation}
should also take care of the SRN regime. In Fig.~\ref{fig:spectra} one can see that least-squares fits, obtained from a Marquadt algorithm with input \eqref{eq:fit_ansatz}, provide excellent approximations in both the SRN and LRN regimes \cite{lsqfit}. The coefficients $\beta$, therefore, provide a measure of the spectral decay rates, which are displayed in the SRN regime as insets in Fig.~\ref{fig:spectra}. The critical exponent $\nu$ is then obtained as the angular coefficient of a linear fit to $\log \beta \times \log (E_J/h)$ and shown to be universal and independent of boundary conditions, number of sites and even dimensionality, as can be seen in more detail in Fig.~\ref{fig:as_and_bs}. In the latter we also display the robustness of our results by comparing between fit parameters obtained from different system sizes. We also note that \eqref{eq:fit_ansatz} will be useful in an analytical approach to thermalization slowing-down and provides, among other things, an analytical expression for the Kolmogorov-Sinai entropy (\emph{e.g.}~\cite{lakshminarayan2011kolmogorov}).

As can be easily seen in the left panels of Fig.~\ref{fig:spectra}, and particularly for the 1d network, the SRN regime is characterized by a LS that is strongly dominated by a small subset of LEs, the majority of the others bending towards zero. Although the rate with which the LEs bend down decreases with increasing dimensionality, a comparison with the panels on the right renders the distinction between LRN and SRN unmistakable, since LEs in the latter remain steadily above zero -- with the exception of the ones associated to the two conserved quantities. This dramatic contrast between SRN and LRN regimes justifies referring to thermalization slowing-down as arising in two distinct universality classes (UCs), depending on which integrable regime is neared. The fundamental distinction between the LRN and SRN UCs is, as previously mentioned, the network range spanned by the nonintegrable perturbation amongst the actions of the integrable limit system \cite{mithun2021fragile,supmat}. As a result, in the LRN class the sites remain strongly coupled as the integrable limit is approached, while decoupling in the SRN class.

It is enlightening to recast the previously described UCs in the language of classical perturbation theory, which cannot be properly applied for the maps studied in \cite{malishava2022lyapunov}. In their integrable limits, both UCs allow for a description in terms of action-angle coordinates, in which the $N$ actions provide a full set of conserved quantities in involution \cite{arnol2013mathematical, de1988hamiltonian}. In the SRN-UC the actions of the unperturbed integrable system are given by the kinetic energy of each uncoupled site, while in the LRN-UC they are the normal modes. Therefore, the actions are spatially extended in the latter, and localized in the former. Once the systems are perturbed, chaos enters the LRN-UC in the form of non-linearly coupled normal modes interacting with each other in an all-to-all fashion, while in the SRN-UC it comes in the form of sparse isolated resonance pockets between weakly interacting nearest-neighbour rotors \cite{livi1987chaotic, livi1987liapunov}. Those pockets are the source of strongest chaos and the main responsible for the magnitude of the mLE in the SRN regime.

Due to the extended nature of perturbed actions in the LRN-UC, any normal mode will be involved in at least one chaotic resonant interaction \cite{chirikov1979universal}, with probability exponentially close to 1 \cite{supmat}. Therefore, in this black and white picture in which a mode is either resonant or not, all LEs should be of the same order of magnitude, confirming that the rescaled LS should approach a stationary curve. At variance, for the SRN-UC a rotor will be involved in a resonance with at least one of its nearest neighbours with a small probability, namely $2d \, E_J/h$ \cite{supmat}. Such sparse resonant pockets have an average distance of $\sim (h/E_J)^{1/d} / d^{1/d}$. Therefore the mLE should show a similar dependence on $E_J/h$ for any dimension, and most of the other LEs should become negligibly small compared to the mLE, confirming the observed exponential decay of the LS.

The above development helps us understand several facets of Figs.~\ref{fig:spectra} and \ref{fig:lambda1s}. Firstly, due to the sites interacting only with their nearest neighbors, it explains why the mLEs in Fig.~\ref{fig:lambda1s} scale according to the same critical exponent in the SRN-UC, predicted analytically to be $2/3$ for the 1d chain \cite{casetti1996riemannian} (see also \cite{parisi1986scaling}) and shown here, perhaps surprisingly, to be the same for all dimensionalities. We note that while \cite{casetti1996riemannian} enters the SRN regime by varying the energy density, we do so by varying the Josephson coupling. The two strategies are not equivalent but can be bijectively mapped, as shown in our Supplemental Material. Indeed, the scaling found by \cite{casetti1996riemannian} in the SRN regime is $h^{-1/6}$, which using Eq.~1 of the Supplemental Material can be easily shown to result in a $h^{-2/3}$ scaling in our case. This has to be contrasted with the LRN-UC, in which all normal modes interact with each other and increasing the system's dimensionality strongly impacts the dynamics, since the number of almost-resonant normal modes increases: $1$ or $2$ for $d=1$ (depending on the boundary conditions), and $\sim N^{d-1}$ for higher dimensions. This explains the smaller slopes of the LRN spectra in Fig.~\ref{fig:lambda1s} as a function of increasing dimensionality -- larger systems are more chaotic and, therefore, the speed at which they stop thermalizing is slower. Interestingly, this is also reflected in our computational simulations, in the sense that to achieve convergence in the LRN regime we need considerably shorter times when dealing with 2 and 3d systems when compared to 1d ones. 

The earlier localization argument also applies to the observed universality in the second critical exponent that characterizes the SRN-UC, denoted as $\nu$ in the insets of Fig.~\ref{fig:spectra}: Since the speed at which exponentially decaying LS curves bend towards zero can only depend on nearest neighbors, the exponent $\nu$ should not depend on dimensionality at all. This is in strong contrast with the LRN-UC, in which the plots in Fig.~\ref{fig:as_and_bs} point to the possible existence of an asymptote that, at present, lies beyond our computational resolving power, but that has been already observed in 1d unitary circuits maps \cite{malishava2022lyapunov}, and indications of which were reported for a short chain of interacting classical spins \cite{constantoudis1997nonlinear}. 

Although both SRN and LRN UCs reflect the fact that the systems stop thermalizing when an integrable limit is reached, the simultaneous presence of large and near-zero LEs in the SRN class is associated to the coexistence of near-conserved quantities and strong chaos -- a characteristic that is completely absent in the LRN class. In the language of KAM theory, one can interpret the SRN road to integrability as the gradual formation of regular low-dimensional submanifolds within chaotic regions in phase space, which continuously become $N$-dimensional tori when the limit is reached. Here, the universality of critical exponents also shows that the speed at which regularity is reconstructed appears remarkably constant. For the LRN pathway, on the other hand, tori are rebuilt as the whole dynamics approaches regularity. If one interprets chaos and integrability as \emph{dynamical phases} \cite{butera1987phase}, the mechanisms for thermalization slowing-down in the SRN and LRN classes are associated to different types of phase transitions.

An interesting future research direction concerns adding disorder. This would allow for an interpolation between different universality classes, since Anderson localization can force the normal modes to localize. An even more intriguing question concerns the corresponding quantum many-body case. We have good reasons to expect that quantizing a classical SRN system will lead to quantum many-body localization, which implicates in the complete destruction of thermalization at a finite distance from the integrable limit. On the other hand, the quantization of a LRN system will probably not destroy thermalization and keep the system in range of the eigenstate thermalization hypothesis.

\medskip
\acknowledgements
GML acknowledges the financial support from the Institute for Basic Science (IBS) in the Republic of Korea through the project IBS-R024-D1. We thank Marco Baldovin, Barbara Dietz, Emanuelle Dalla Torre, Yagmur Kati, Dario Rosa, Steven Tomsovic, Angelo Vulpiani, Weihua Zhang and Tilen \v{C}ade\v{z} for fruitful discussions, and Robert McLachlan for pointing out the advantages described in \cite{mclachlan2001kinds}.

\bibliography{bib}

\end{document}


\title{Supplemental Material for ``Thermalization slowing down in multidimensional Josephson junction networks''}
\author{Gabriel M.~Lando}
\author{Sergej Flach}
\affiliation{Center for Theoretical Physics of Complex Systems, Institute for Basic Science, Daejeon 34126, Korea}

\maketitle

The objective of this supplement is to provide a comprehensive account of all numerical and analytical details omitted from the main text, together with stringent tests for the convergence of our results. In the following we use the conventions adopted in the main text, namely that $H$ and $\Phi$ are the Hamiltonian and its flow and $\mathbf{z}=(\mathbf{p},\mathbf{q})$ denotes a $2N$-dimensional phase space vector. 1-, 2- and 3-dimensional networks are denoted by writing their number of sites as a power of dimensionality, and here the $N=12^2$ network is used for the most delicate numerical tests. We will also refer to rescaled Lyapunov spectra as LS, dropping the ``rescaled''. 

\section{Rescaling transformations}\label{sec:rescaling}
The relevant control parameter of our system is $E_J/h$. Let $\mathbf{P}(t)$ and $\mathbf{Q}(t)$ be the projections of the flow for some value of $E_J/h$ and initial condition $(\mathbf{p}, \mathbf{q})$ into momentum and position subspaces, \emph{i.e.}~$\mathbf{P} = \pi_1 \circ \Phi$, $\mathbf{Q} = \pi_2 \circ \Phi$, where $\pi_1(\mathbf{p},\mathbf{q}) = \mathbf{p}$ and $\pi_2(\mathbf{p},\mathbf{q}) = \mathbf{q}$ are the canonical projections. Any $\mathbf{Q}(t), \, \mathbf{P}(t)$ can be transformed into $\mathbf{\tilde{Q}}(\tau)=\mathbf{Q}(\mu t)$ and $\mathbf{\tilde{P}}(\tau)=\mathbf{P}(\mu t)/\mu$, $\tau=\mu t$ and $\tilde{E}_J = E_J/\mu^2$ for an arbitrary real and positive choice of $\mu$. The transformed trajectory evolves at an energy density $\tilde{h}=h/\mu^2$ such that $\tilde{E}_J/\tilde{h} = E_J/h$. For any practical purposes we perform computations for $E_J \leq h$ at $h=1$ and running time $t$. Instead for $E_J > h$ we fix $E_J=1$ and rescale time accordingly. This is done in order to avoid the numerical issues of dealing with large $h$ or $E_J$ values, which cause either the kinetic of potential terms to blow up. The above transformations avoid this problem by allowing us to deal exclusively with decreasing either $E_J$ or $h$ while holding the other parameter fixed. For the LS, for example, we connect the different regimes with
\begin{equation}\label{eq:mapping}
    \Lambda_{E_J}(z) = \frac{\Lambda_h(z)}{\sqrt{h}} \, \quad ,
\end{equation}
where the subscript denotes which quantity is being varied.   

\section{Networks}\label{sec:networks}

In the LRN regime $h/E_J \ll 1$ we can assume $E_J=1$ and $|p|,|q| \ll 1$. At the integrable limit we expand the cosine interaction potential to 2nd order, $\cos(x) \approx 1-x^2/2$. In proximity to that limit the leading order nonintegrable perturbation results from the next quartic term in the Taylor expansion of the potential, namely $\cos(x) \approx 1-x^2/2 + x^4/24$. For simplicity we stay in dimension 1 (extension to higher dimensions is straightforward). The resulting Hamiltonian is a variant of the celebrated $\beta$-Fermi-Pasta-Ulam-Tsingou model
\begin{equation}
    H = \sum_l \frac{p_l^2}{2} + \frac{E_J}{2} (q_l - q_{l-1})^2
    -\frac{E_J}{24} (q_l-q_{l-1})^4
    \;. 
    \label{FPUT}
\end{equation}
We use the canonical transformation to normal mode momenta and coordinates 
$\{P_k,Q_k\}$
\begin{equation}
    \begin{split}
        \left(
        \begin{array}{c}
        P_k\\
        Q_k\\
        \end{array}
        \right)
        &= \sqrt{\frac{2}{N+1}}\sum_{n=1}^{N}
        \left(
        \begin{array}{c}
        p_n\\
        q_n\\
        \end{array}
        \right)
         \sin \bigg(\frac{\pi n k}{N+1}\bigg)
    \end{split}
    \label{eq:Fourier coordinates}
\end{equation}
for $k=1,\dots,N$. This transformation diagonalizes the integrable quadratic Hamiltonian part  $H_0 = \sum_{k=1}^N E_k$ in Eq.(\ref{FPUT}), where the normal mode energies $E_k$ are 
\begin{equation}
    E_k = \frac{P_k^2 + \Omega_k^2 Q_k^2}{2}\ , \qquad \Omega_k = 2\sqrt{E_J}\sin\left(\frac{\pi k }{2(N+1)}\right)\;.
    \label{eq:modes_energy_KG}
\end{equation}
%
The equations of motion in the normal mode coordinates \eqref{eq:Fourier coordinates} then read 
\begin{align}
    \ddot{Q}_k + \Omega_k^2  Q_k =  \frac{E_J}{12(N+1)}  &\sum_{l_1,l_2,l_3}  \Omega_k \Omega_{l_1} \Omega_{l_2} \Omega_{l_3} A_{k,l_1,l_2,l_3} \notag \\
    &\quad \times Q_{l_1} Q_{l_2} Q_{l_3} 
\label{eq:KG_NM_low_e}
\end{align}
where
\begin{equation}
    \begin{split}
    A_{k,l_1,l_2,l_3} &= 
    \delta_{ k - l_1 + l_2 - l_3 , 0}  + \delta_{k - l_1 - l_2 + l_3,0}  \\
    &\quad -  \delta_{k+ l_1+l_2 - l_3,0}  -  \delta_{k + l_1 - l_2  + l_3 , 0}
    \end{split}
    \label{eq:KG_exp_coeff1}
\end{equation}
represents the coupling between the Fourier coordinates $Q_k$. Using the canonical transformation  
\begin{equation}
    \label{eq:can_transf_KG_e0}
    Q_k = \sqrt{2 J_k} \sin \theta_k  \qquad P_k = \Omega_k \sqrt{2 J_k} \cos \theta_k 
\end{equation}
it follows that
\begin{equation}
    \label{eq:KG_action_angle}
    \begin{split}
    &\qquad  \dot{J}_k 
     =  \sum_{l_1,l_2,l_3} \mathcal{A}_{k,l_1,l_2,l_3}  \sqrt{J_k J_{l_1} J_{l_2}  J_{l_3}} 
    \end{split}
\end{equation}
where the coefficients $\mathcal{A}_{k,l_1,l_2,l_3} $ depend on the angles $\{ \theta_k\}_k$:
\begin{equation}
    \label{eq:KG_action_angle_coeff}
    \begin{split}
    &\mathcal{A}_{k,l_1,l_2,l_3} 
     = \frac{E_J}{6} \frac{ A_{k,l_1,l_2,l_3} }{2(N+1)} 
     \Omega_{l_1} \Omega_{l_2} \Omega_{l_3}
     \cos \theta_k \sin \theta_{l_1} \sin \theta_{l_2} \sin \theta_{l_3} \ .
    \end{split}
\end{equation}
Therefore each action $J_k$ is interacting with all other actions via the proliferating number of $\sim N^2$ nonlinear terms on the RHS of \eqref{eq:KG_action_angle_coeff}. This constitutes the LRN imposed by the nonintegrable perturbation onto the set of actions, and the number of terms on the RHS readily generalizes to $N^2d$ for higher dimensions.

In the SRN regime $E_J/h \ll 1$ the integrable limit model is the case of free rotors $E_J=0$. The coordinates $p_l=J_l$ and $q_l=\theta_l$ are already the proper action-angle choice. The nonintegrable perturbation is the entire Josephson coupling potential term in the original Hamiltonian, therefore imposing nearest neighbour couplings between the actions. This constitutes the SRN, where the number of actions a given reference action is interacting with is finite, and not scaling with the system size. The corresponding trivial metric allows to introduce a distance between actions distinguishing large from small distances.

\section{Resonance probabilities}

In the LRN regime a normal mode interacts with $N^{2d}$ additive quadruplets of other modes (see Sec.~\ref{sec:networks}). The probability $p$ of resonance for one quadruplet is obtained from first-order perturbation theory in the nonintegrable perturbation $E_J(q_{\sigma_1}-q_{\sigma_2})^4$ and is easily evaluated to be $p\sim (h/E_J)^2$. The probability for one mode to be resonant with at least one quadruplet is then obtained from the complementary probability not to be in resonance with any of them, which reads $p_R = 1 - (1-p)^{N^2} \approx 1- e^{-pN^2}$. For macroscopic systems it follows that $p_R=1$.

In the SRN regime the first-order perturbation correction to a free rotor dynamics $p_{\sigma_1}^{(0)}=\mathrm{const}$ reads
$p_{\sigma_1}^{(1)} \sim E_J/(p_{\sigma_1}^{(0)}-p_{\sigma_2}^{(0)})$. The probability to be in resonance is then $p_R \sim 2d (E_J/h)$. These sparse resonances are at an average distance $p_R^{-1/d}$ from each other, with this distance increasing upon approaching the integrable limit.

\section{Sampling of initial conditions}\label{sec:sampling}

For all dimensionalities considered in the main text, the total momentum of the system is conserved. Without loss of generality we sample initial conditions according to the following recipe:

\begin{enumerate}
    \item Set $\mathbf{q}(0) = 0$;
    \item Sample the components of the initial momentum according to a $\chi_2$ distribution for the 1- and 2-dimensional models and $\chi_3$ for the 3-dimensional one;
    \item Transform each component of momentum to $p_i(0) \longmapsto p_i(0) - \langle \mathbf{p}(0) \rangle$, where $\langle \mathbf{p}(0) \rangle$ is the mean initial momentum.
\end{enumerate}
$\chi_s$ is the distribution with probability density function
\begin{equation}
    f(x;s) = 
    \begin{cases}
        \dfrac{x^{s-1} e^{x^2/2}}{2^{s/2 - 1} \Gamma(s/2)} &\quad x \geq 0 \\
        0 &\quad \text{otherwise}
    \end{cases} 
    \quad ,
\end{equation}
from which the Maxwell-Boltzmann distribution is recovered for $s=3$. The procedure above results in zero initial momentum, which should be approximately conserved by symplectic integration alongside the energy. The energy density of the system is then fixed and equal to
\begin{equation}
    h = \frac{1}{N} \sum_{i=1}^N \frac{p_i(0)^2}{2} \, ,    
\end{equation}
since $\mathbf{q}(0) = 0$. Through a further rescaling of the initial momentum as $\mathbf{p}(0) \longmapsto \mathbf{p}(0)/\sqrt{k}$ one is effectively dividing the energy density by $k$, such that by increasing $k$ while keeping $E_J$ fixed we can access the LRN regime. The SRN regime, on the other hand, is accessed by keeping $k=1$ and decreasing $E_J$, as discussed in Sec.~\ref{sec:rescaling}.

For dynamics in the tangent space, the initial monodromy matrix to be propagated is chosen as
\begin{equation}
    \mathcal{M}(0) = 
    \begin{pmatrix}
        I & 0 \\
        0 & \widetilde{I}
    \end{pmatrix} \,, \quad
    I_{ij} = \delta_{i,j} \,, \quad \widetilde{I}_{ij} = \delta_{i,2N-j+1} \quad ,
\end{equation}
which departs from the established choice of just propagating the identity matrix. Nevertheless, this simple permutation of rows guarantees that the sum of positive and negative exponents is conserved in time and machine-equal to zero -- an important property of time-independent Hamiltonian systems \cite{mclachlan2001kinds}. 

\section{Conserved quantities}

\begin{figure*}
   \centering
    \begin{tabular}{cc}
        \includegraphics[width=0.5\linewidth]{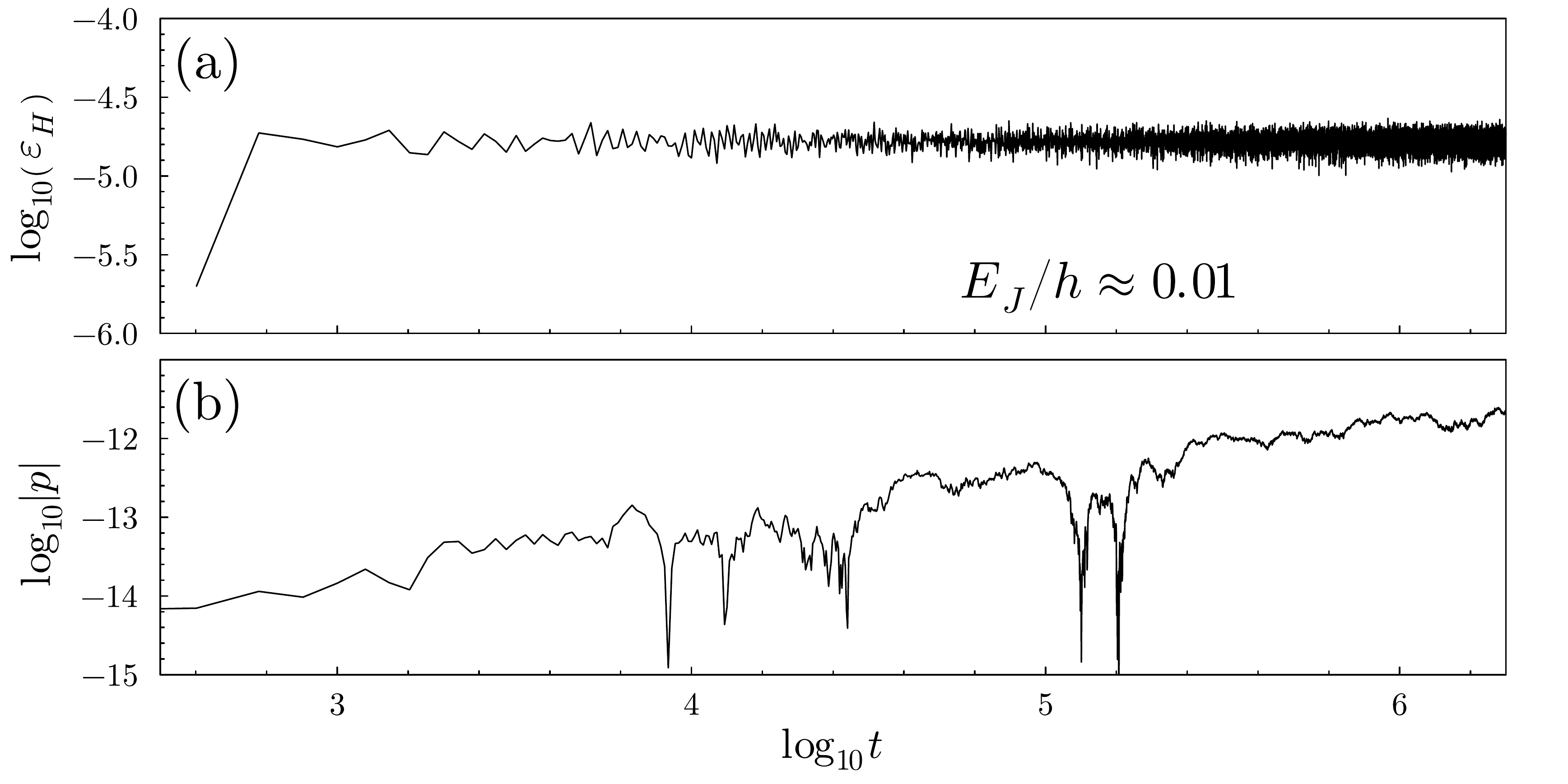} & \includegraphics[width=0.5\linewidth]{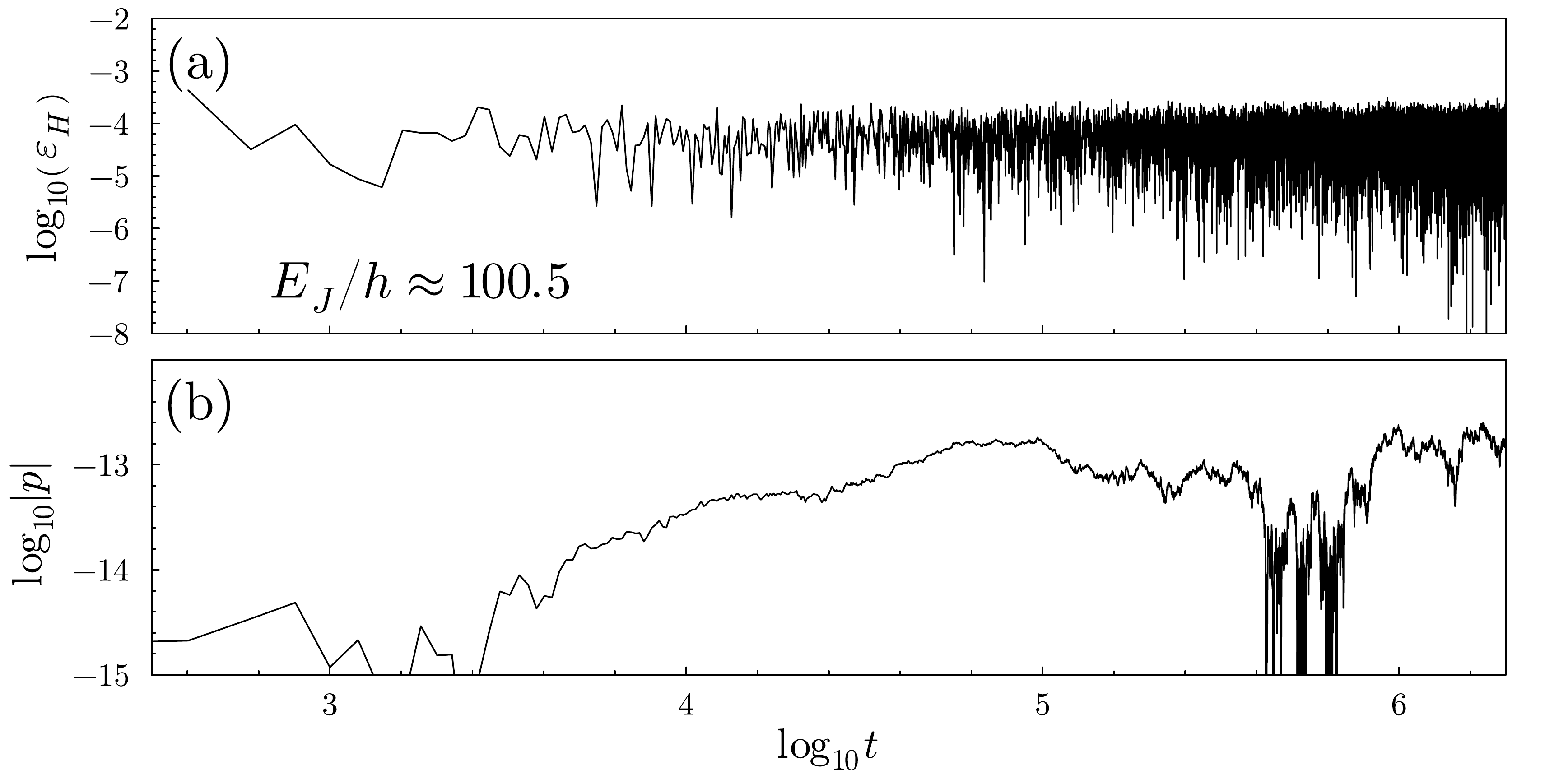} \\
        \includegraphics[width=0.5\linewidth]{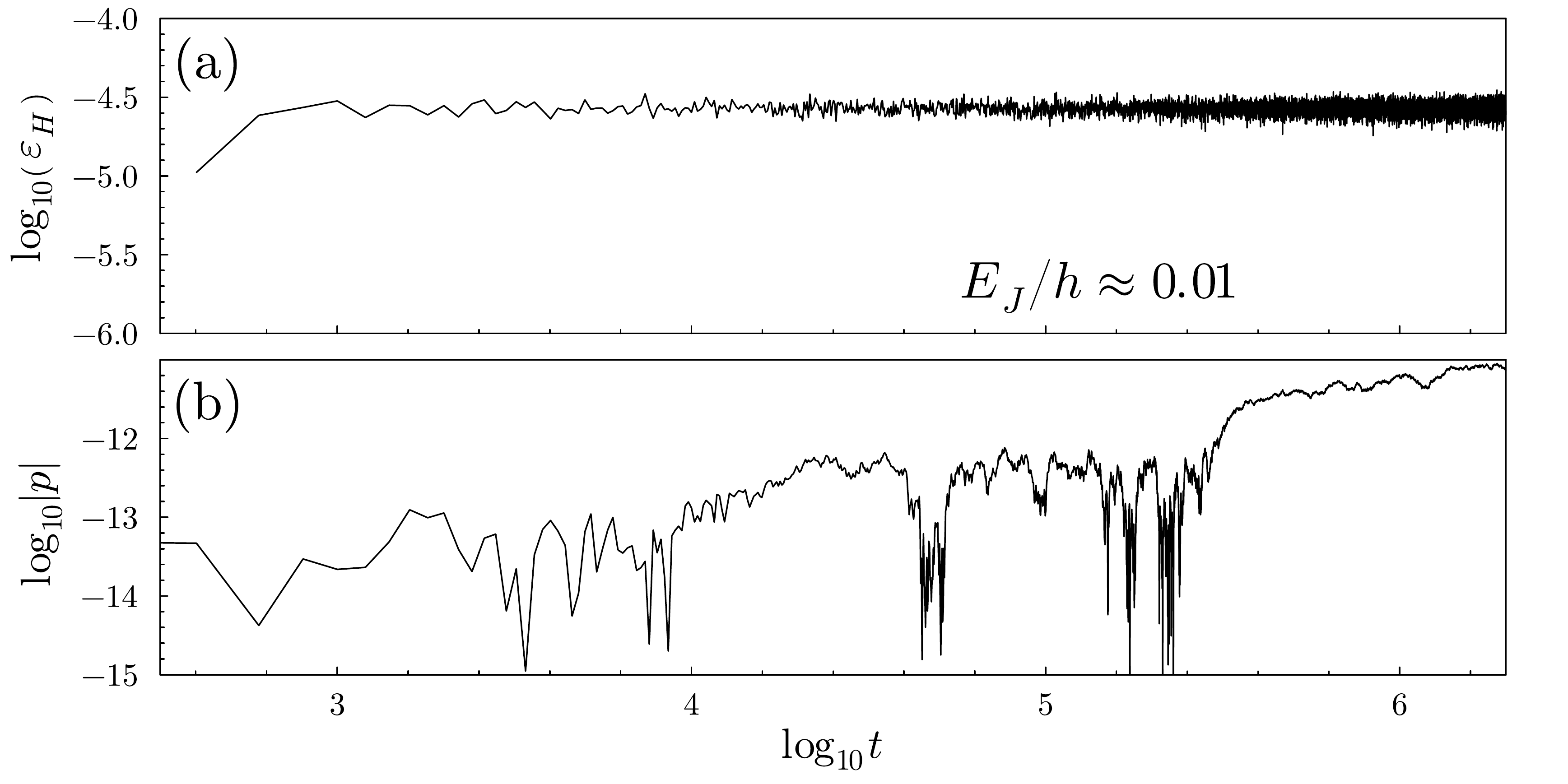} & \includegraphics[width=0.5\linewidth]{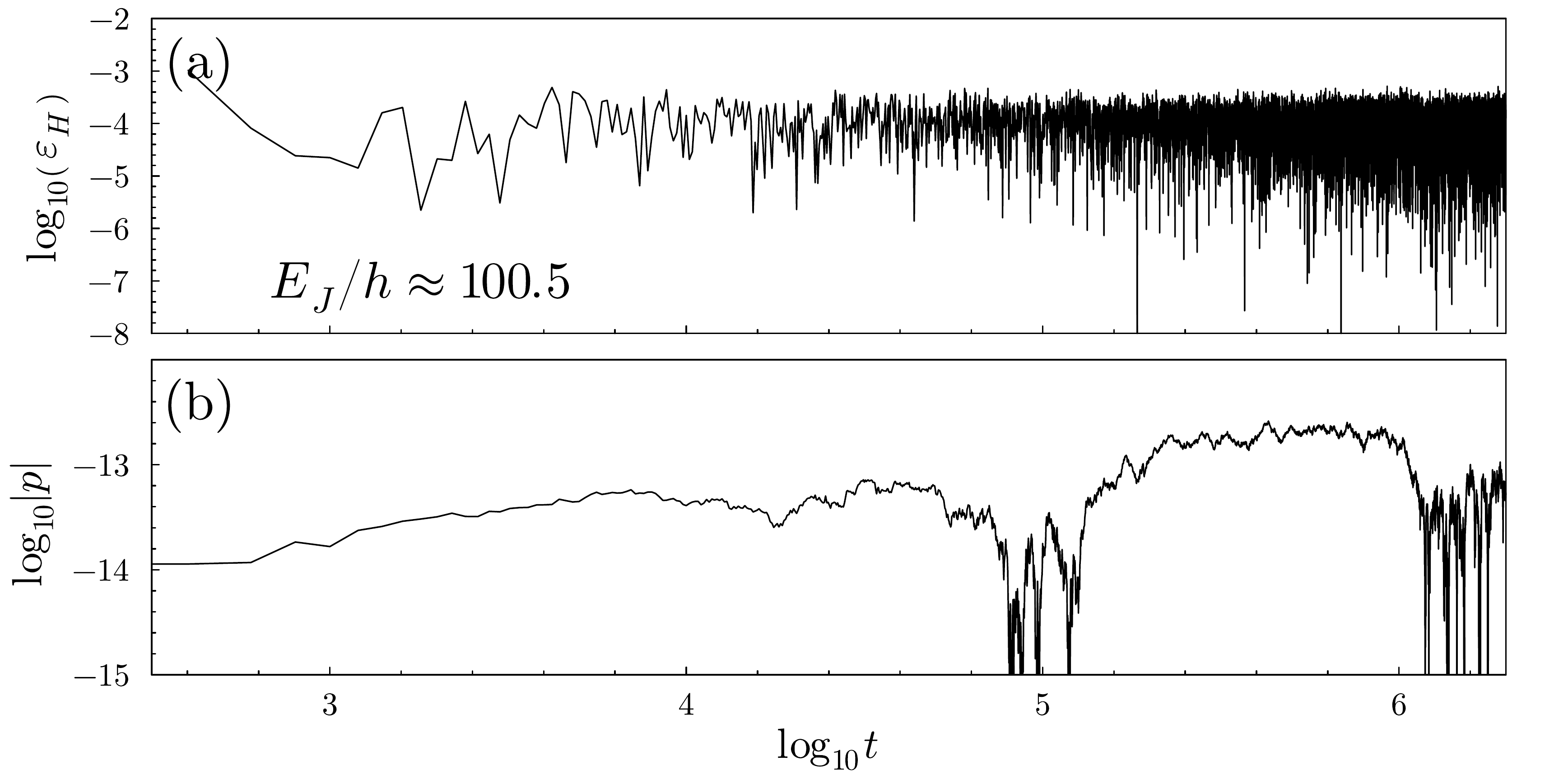} \\
        \includegraphics[width=0.5\linewidth]{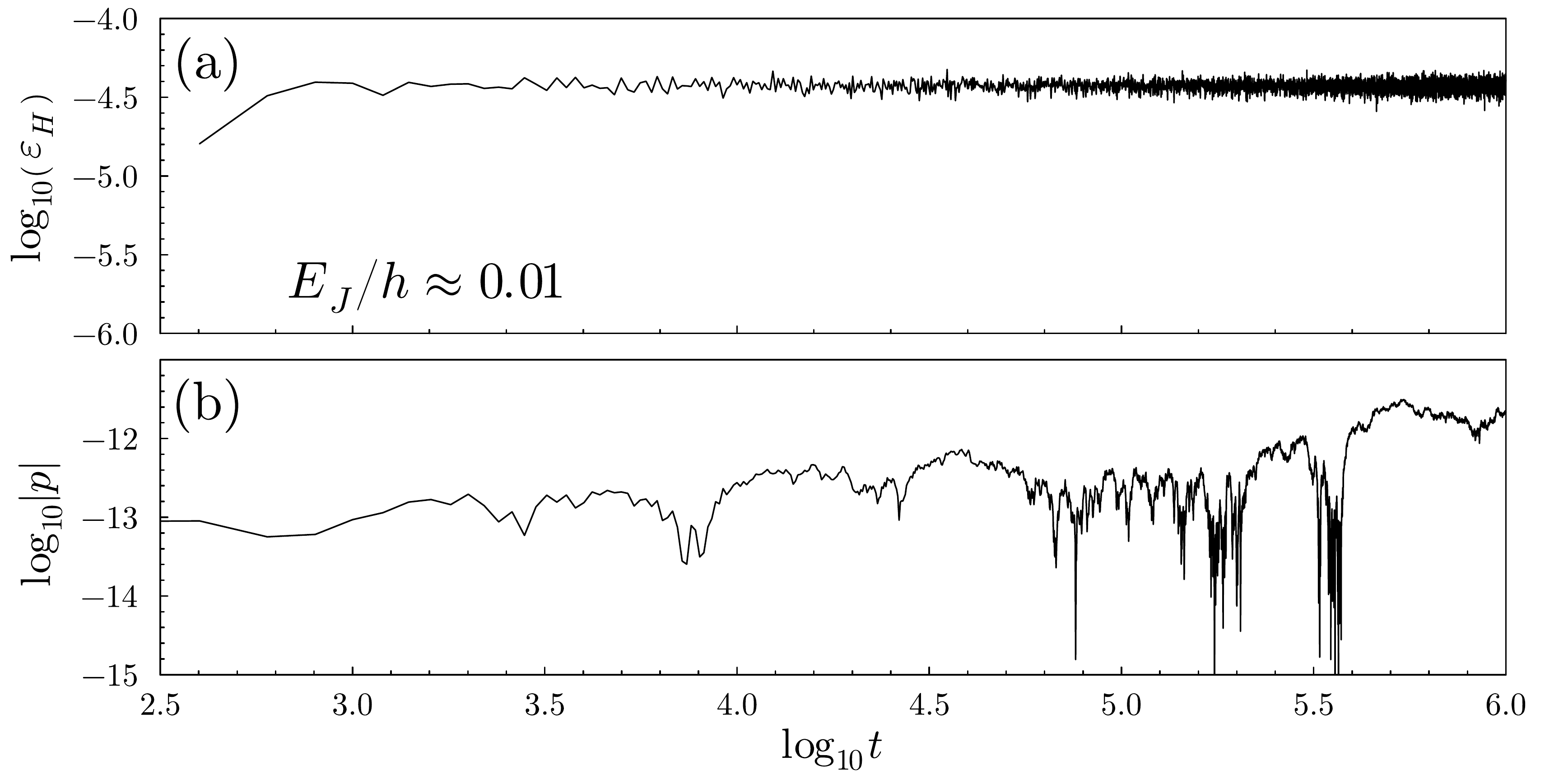} & \includegraphics[width=0.5\linewidth]{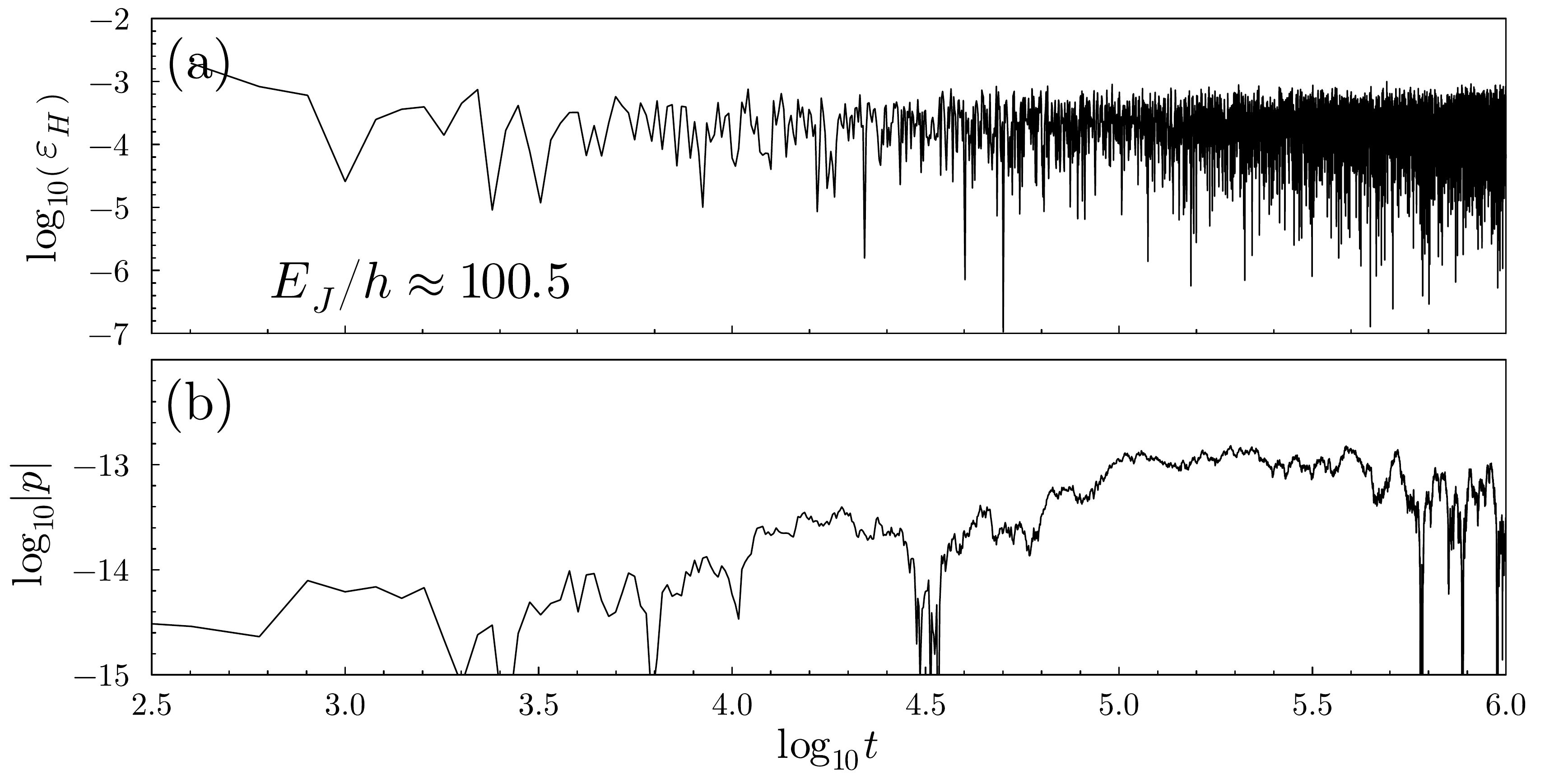} \\
    \end{tabular}
    \caption{(Left panels) Relative energy (a) and total momentum (b) errors for $N=140$ (upper panels), $N=12^2$ (middle panels) and $N=5^3$ (bottom panels) in the SRN regime. The final times are the same as the ones used to calculate the spectra. All errors remain bounded below $10^{-4}$, but become more stable as the dimensionality is increased. (Right panels) Same but for the LRN regime.}
    \label{fig:errors} 
\end{figure*} 

Symplectic integrators (SIs), which are the method we chose to compute the flow in phase space and tangent space, come in a variety of flavours with respect to accuracy and stability \cite{yoshida1990construction, mclachlan1992accuracy}. Typically, these integrators are chosen to be of high-order, with their large number of iterations per step compensated by choosing big time steps \cite{kinoshita1991symplectic, choi2021high, laskar2001high, mogavero2023timescales}. However, the QR-decomposition method used to compute the LS works best with small time steps \cite{mclachlan1992accuracy}, which would render high-order integrators prohibitively slow. We have indeed consistently observed that LS converge more smoothly when employing a low-order integrator with a smaller time step than a high-order one with a large time step. This favors the choice of a low-order integrator, despite the increase in computational cost. In order to avoid triggering numerical degeneracies in $\mathcal{M}$ as it evolves in time, we have also chosen to apply the QR-decomposition method at every time step instead of waiting, which slows down computations but is necessary due to several LEs in the spectra being extremely small and easily lost \cite{benettin1980lyapunov, geist1990comparison}.

\begin{figure*}[t!]
    \centering
    \includegraphics[width=0.9\linewidth]{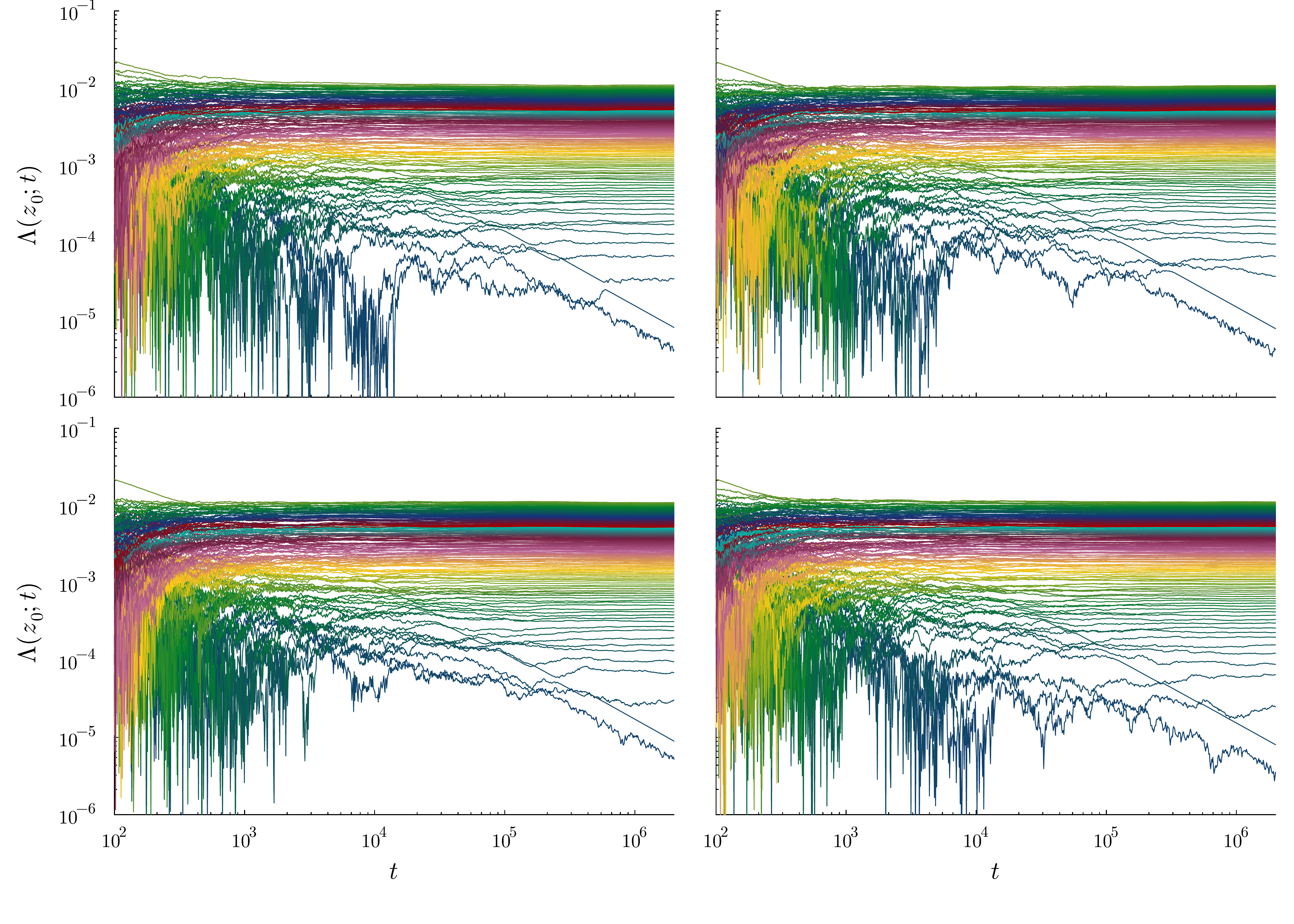}
    \caption{Unrescaled LS for the $N=12^2$ network using four different initial conditions, $z_0$, as a function of time, with $t_f = 2 \times 10^6$. Here, we are far from both the LRN and the SRN regimes, with $\log_{10}(E_J/h) \approx 0.5$.}
    \label{fig:LS}
\end{figure*}

\begin{figure*}
   \centering
    \begin{tabular}{c}
        \includegraphics[width=0.9\linewidth]{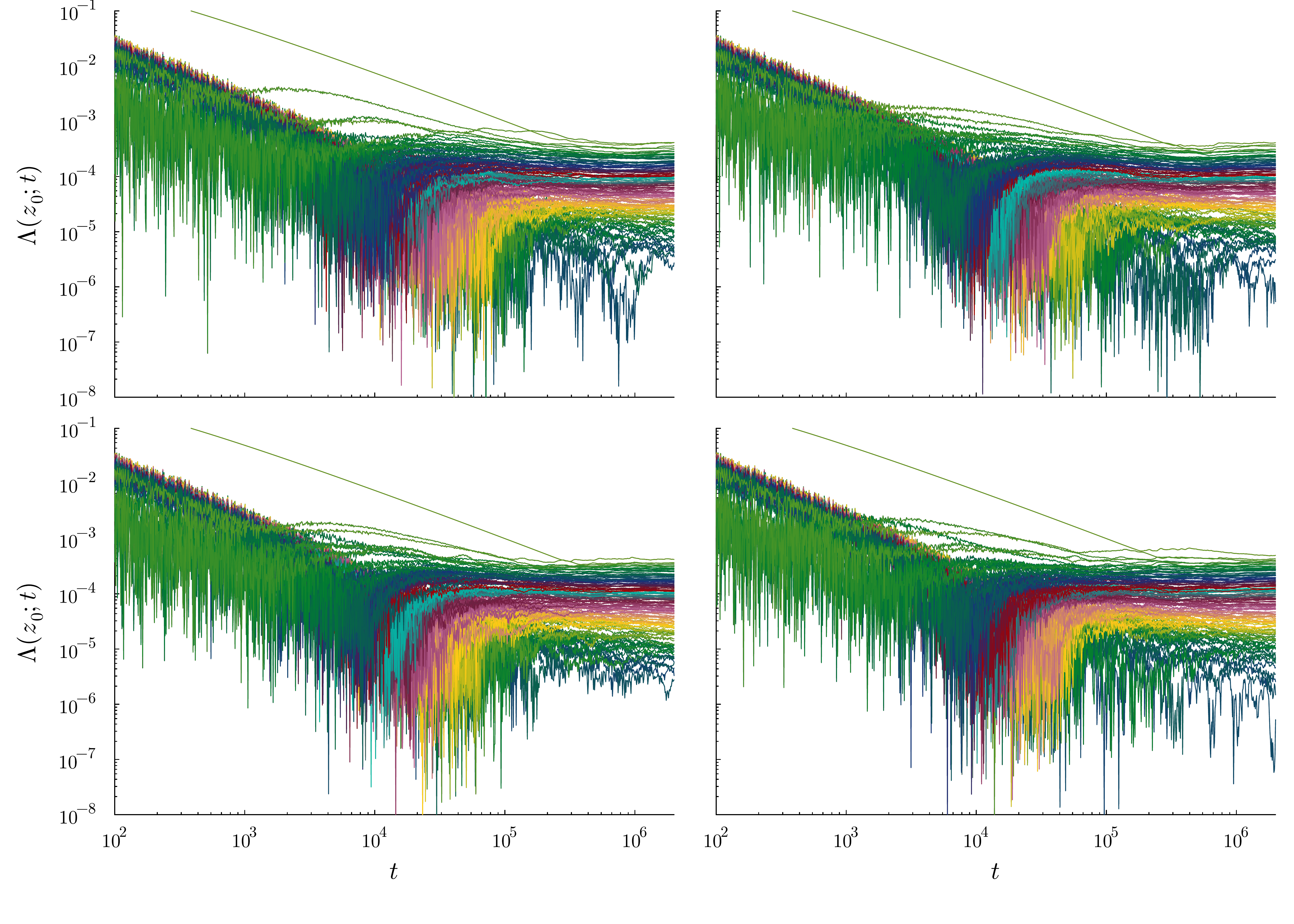} \\ 
        \includegraphics[width=0.9\linewidth]{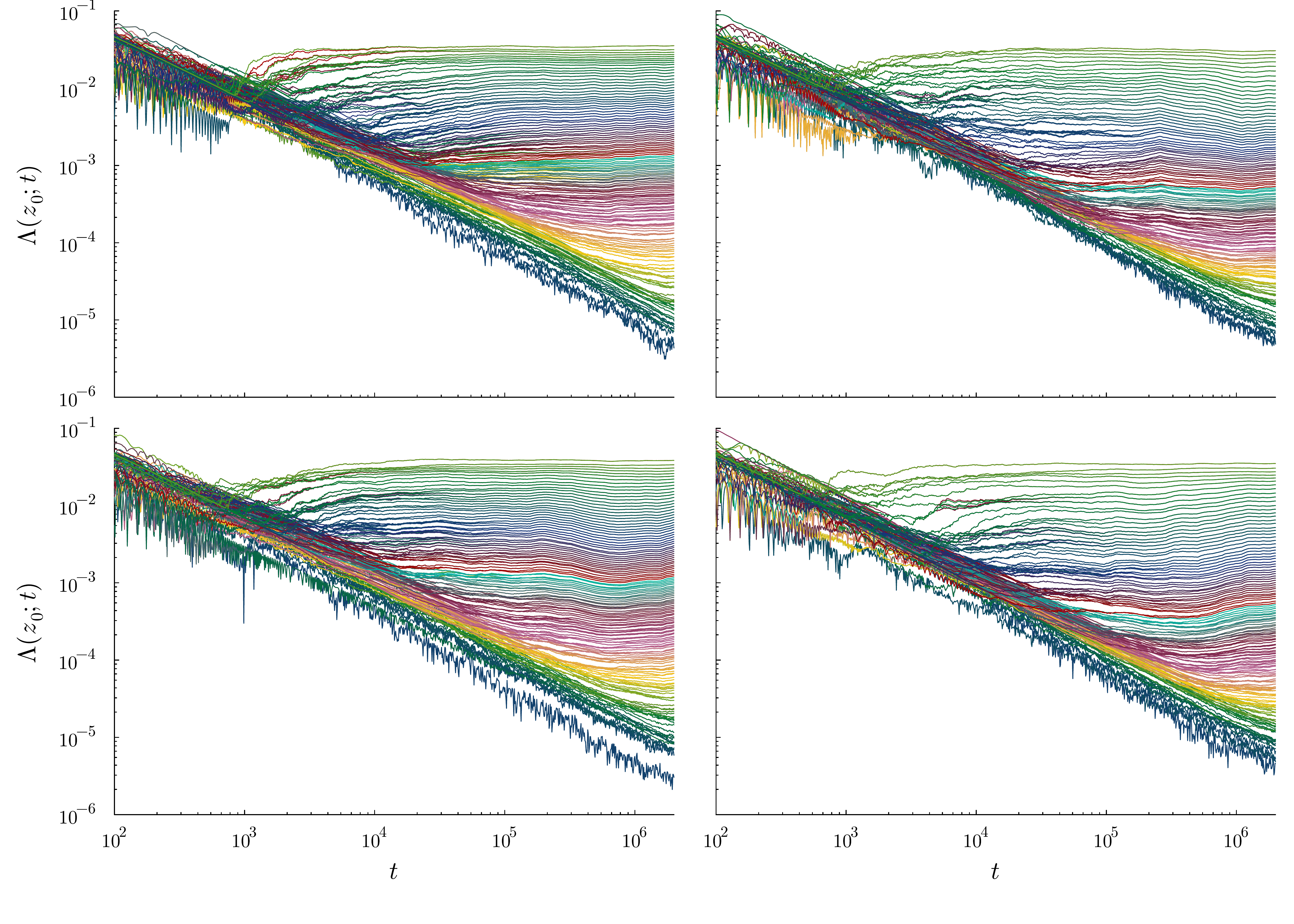}
    \end{tabular}
    \caption{Same as Fig.~\ref{fig:LS}, but now in the deep LRN regime with $\log_{10}(E_J/h) \approx 2.2$ (top 4 panels) and in the deep SRN regime with $\log_{10}(E_J/h) \approx -2.2$ (bottom 4 panels).}
    \label{fig:LRN_SRN} 
\end{figure*} 

Nevertheless, SIs do not exactly preserve conserved quantities, and their accuracy is usually measured in terms of ``drifts'' in things algebraically known to be constant. In strongly chaotic systems like the ones we deal with, this is in fact the only way one can check the quality of an integrator, since the trajectories for one particular choice of $\text{d}t$ will be completely different than for another due to exponential sensitivity -- thus, trajectory errors are essentially useless. Note, however, that LS computed with different $\text{d}t$s should be very close to each other due to the shadowing lemma \cite{palmer1988exponential} (see Sec.~\ref{sec:robustness}). 

In the left panels of Fig.~\ref{fig:errors} we display the drifts (\textit{i.e.}~the errors) in the deep SRN regime for all dimensionalities discussed in the main paper, and three choices of $N$ (the number of sites). The [relative] energy drift is defined as
\begin{equation}
    \varepsilon_H = \left\vert 1 - \frac{(H\circ\Phi)(\mathbf{z};t)}{H(\mathbf{z})} \right\vert \quad, 
\end{equation}
while the momentum error is just the norm of the momentum vector, which due to our initial conditions should be conserved and equal to zero. Drifts are shown as a function of time, starting from $10^2$ up to the final times used for computing the LS. The same is done in the right panel of Fig.~\ref{fig:errors}, but this time for the LRN regime. Note that the errors are larger in LRN than in SRN, being bounded by $10^{-3}$ and $10^{-4}$, respectively, and also less stable for the former than the latter. This is an indication that the LRN regime is numerically harder to deal with than the SRN one, which can also be noted from the fact that their LEs are much smaller than their SRN counterparts. 

We only display the errors near integrability because in the intermediate regime, say $| \log_{10}(E_J/h) | < 0.5$, achieving convergence in the spectra is very easy and, in principle, one does not even need to go to very long times -- although, for consistency, we did use the same final times to cover the whole $E_J/h$ axis. 

\section{Convergence of LS}

We now move on to quantifying the errors in the LS as a function of time. The easiest regime to deal with is surely $| \log_{10}(E_J/h) | < 0.5$, as stated earlier. As as example, in Fig.~\ref{fig:LS} we display the time-dependent LS for 4 different initial conditions, obtained following the sampling procedure described in Sec.~\ref{sec:sampling}, using the $N=12^2$ system as an example. Since there are 144 LEs in the LS, it is essentially impossible to see them simultaneously, such that we color them based on their values to facilitate visualization. This figure shows several important trends: 
\begin{enumerate}
    \item We are able to resolve the whole spectrum, including the two zeros, which are the two lowest LEs decaying as $1/t$;
    \item All runs produce essentially the same final spectrum at $t=t_f$;
    \item Almost all LEs, including the small non-zero ones, have remained stable since $t \approx 10^5$.
\end{enumerate}

\begin{figure*}
    \centering
    \includegraphics[width=\linewidth]{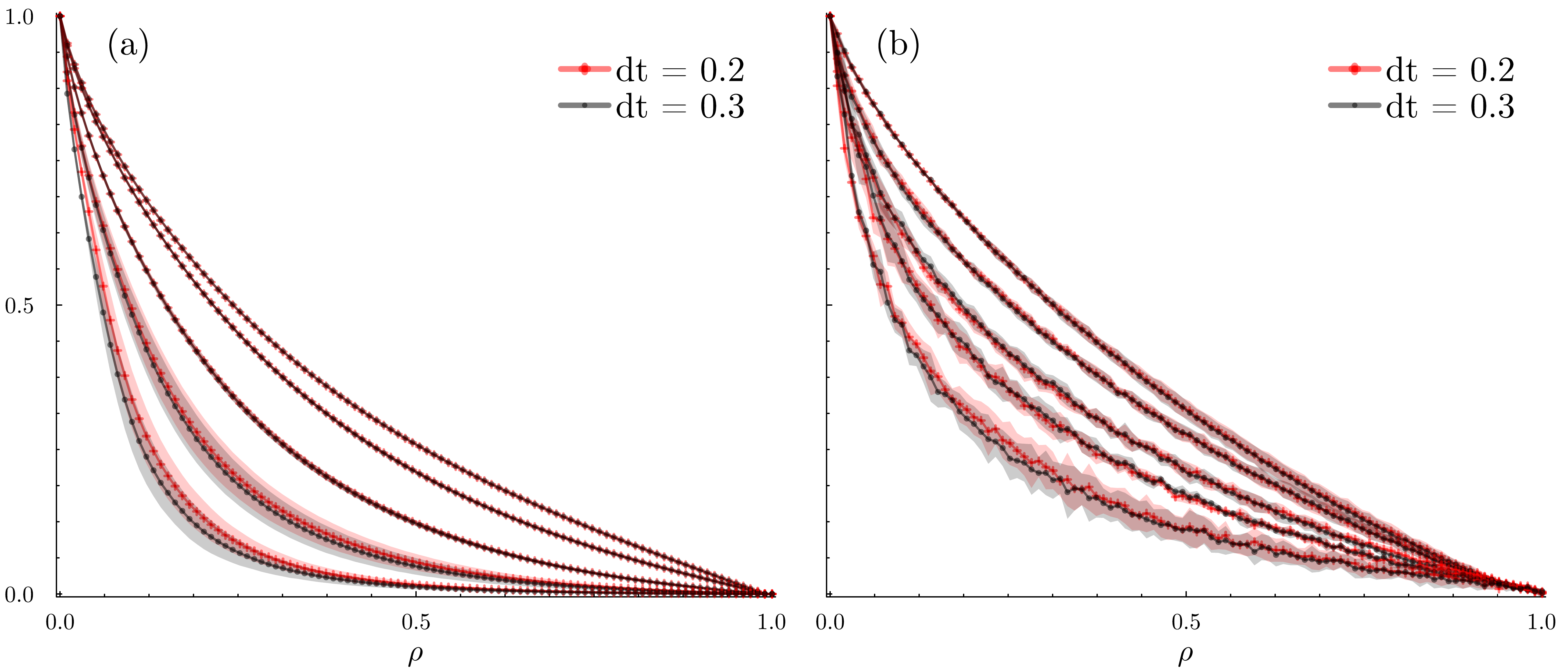}
    \caption{Qualitative comparison between LS obtained with two different step sizes for $N=12^2$. Panel (a) shows LS in the SRN regime, while (b) is similar but in the LRN regime.}
    \label{fig:convergence_spectra}
\end{figure*}

In the above, item (1) is a proof that we have achieved thermalization/equipartition in our numerical simulations for fixed initial conditions, but (2) shows that all initial conditions lead to the same result -- \emph{i.e.}~we have achieved true numerical ergodicity. Point (3) shows that LS are properly converged, and one must not be fooled by the logarithm scales in the time axis: Convergence between $10^5$ and $2\times10^6$ means that the LS were stable for 95\% of the simulation time. 

What we see in Fig.~\ref{fig:LS} is the perfect scenario one can achieve when computing LS. As we approach an integrable limit, the situation will be different and the LS will display both finite-time and finite-size effects. In Fig.~\ref{fig:LRN_SRN} we present the equivalent of Fig.~\ref{fig:LS}, but now deep in the LRN and SRN regimes with $\log_{10}(E_J/h) \approx 2.2$ and $\log_{10}(E_J/h) \approx -2.2$, respectively. In the LRN it is clear we were able to resolve the whole finite component of the LS, although to resolve the zeros we would need even longer times. Nevertheless, the finite LEs have clearly remained stable for the same interval as the ones in Fig.~\ref{fig:LS}, and all runs show very little dependence on the initial condition. In the SRN case, however, one can see that there are visible differences between the different runs due to the some weak dependence on initial conditions, which is a manifestation of finite size effects. It is also clear that we are not able to resolve the full finite component of the LS, since we know that only two exponents truly decay as $1/t$, while in the figures we can see that a handful of them are still behaving as numerical zeros and have not been resolved. We consider that being able to resolve around 97\% of the LEs is a remarkable achievement in terms of numerical and algorithmic stability, and the exponents that we are unable to catch are essentially invisible in the final spectrum -- they are around ten thousand times smaller than the mLE. 

Resolving the full LS becomes a much harder task in the 1d systems due to the reasons discussed in the main paper, such that we are able to resolve around 30 or 40 out of 100 exponents for the $N=100$ case. However, almost all of the finite LEs in the deep SRN regime are several orders of magnitude smaller than the mLE, such that the resolution we achieve is more than enough to characterize the 1d system as well. As a further confirmation, we also employ two system sizes in all of our computations, such that very strict numerical convergence is required given that the results behave as simarly as seen in Figs.~2 and 3 in the main paper.

Now that we have shown how the LS behave as a function of time and initial condition, we can define how we quantify errors in the spectra. For each initial condition, we define the final LS corresponding to it as being the time-average of the time-dependent LS for half the simulation time. Then, the final LS is given by the average of all LS computed with the same parameters, but different initial conditions, and the error is defined as their standard deviation. This is done for every value of $E_J/h$. We run 7 different initial conditions for the 1d systems because they are less stable, but for the 2d and 3d systems 3 or 4 runs are enough for accurate error estimates. The time needed to obtain a single LS varies between three and fifteen days in core-hours. From then on, we perform all the procedures (spectral fits, critical exponent fits, etc) using the mean spectra. The uncertainty in the critical exponents is obtained as the margin error at 5\% significance \cite{lsqfit}, and the two critical exponents obtained for different system sizes agree within their error bounds. 

\section{Numerical robustness}\label{sec:robustness}

Stiffness is a fundamental effect to be taken into account when employing numerical methods to solve ordinary differential equations. Loosely speaking, a system is stiff if the time-step required to enforce error boundedness tends to zero, despite the solution curve being well-behaved (\emph{i.e.}~smooth or $C^r$ for large $r$) \cite{wanner1996solving}. Although a precise definition of stiffness is lacking, its impact is easily identifiable when employing adaptive step-size methods due to anomalous behavior in the step sizes. Naturally, this behavior will quantitatively depend on the step-size controller, but stiff methods are usually assumed to trigger anomalies independently of how the step sizes as chosen. 

\begin{figure}
    \centering
    \includegraphics[width=\linewidth]{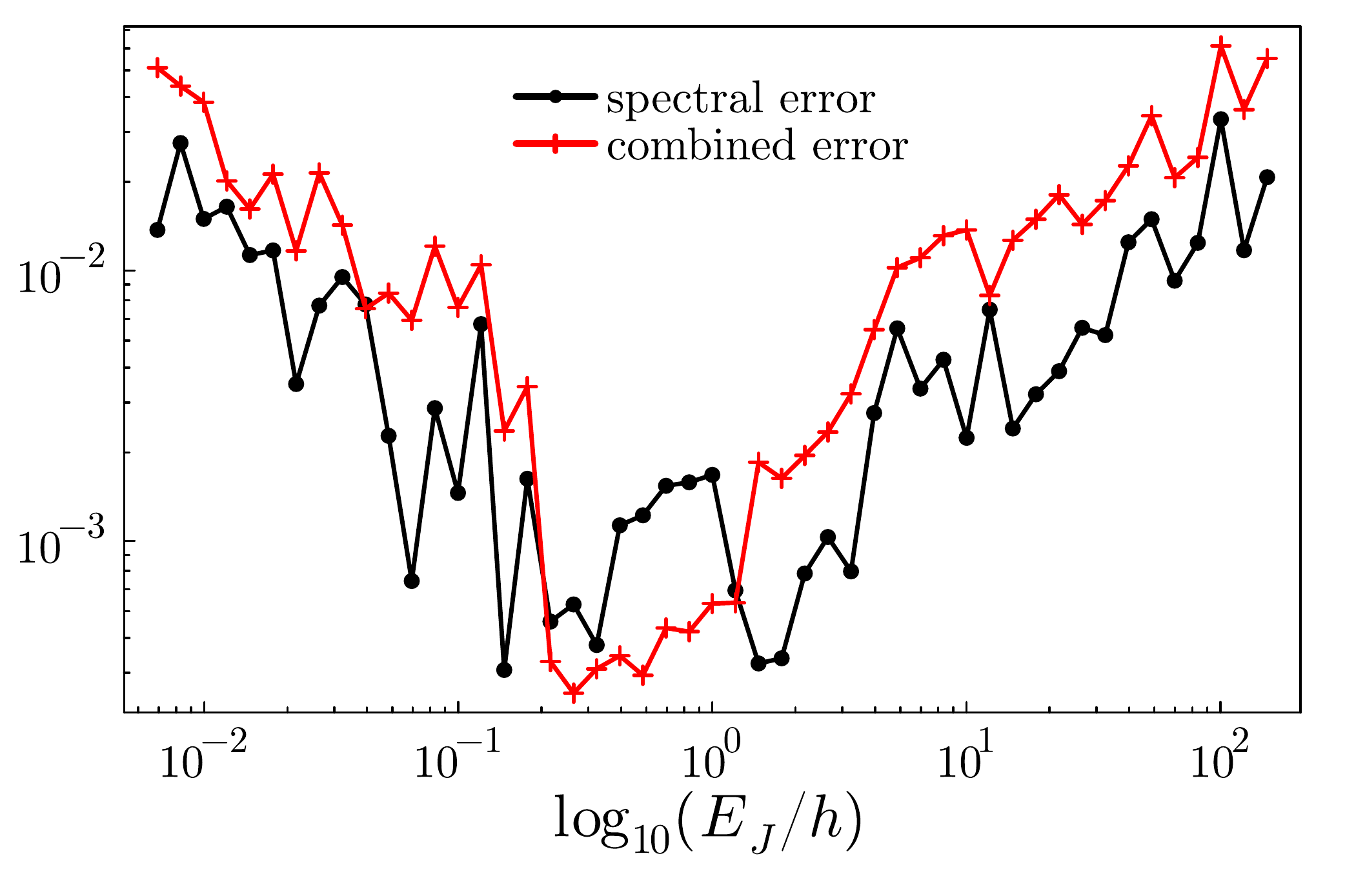}
    \caption{Spectral error \eqref{eq:se} versus combined error \eqref{eq:ce} as a function of $E_J/h$.}
    \label{fig:convergence_error}
\end{figure}
\begin{figure}[t!]
    \centering
    \includegraphics[width=\linewidth]{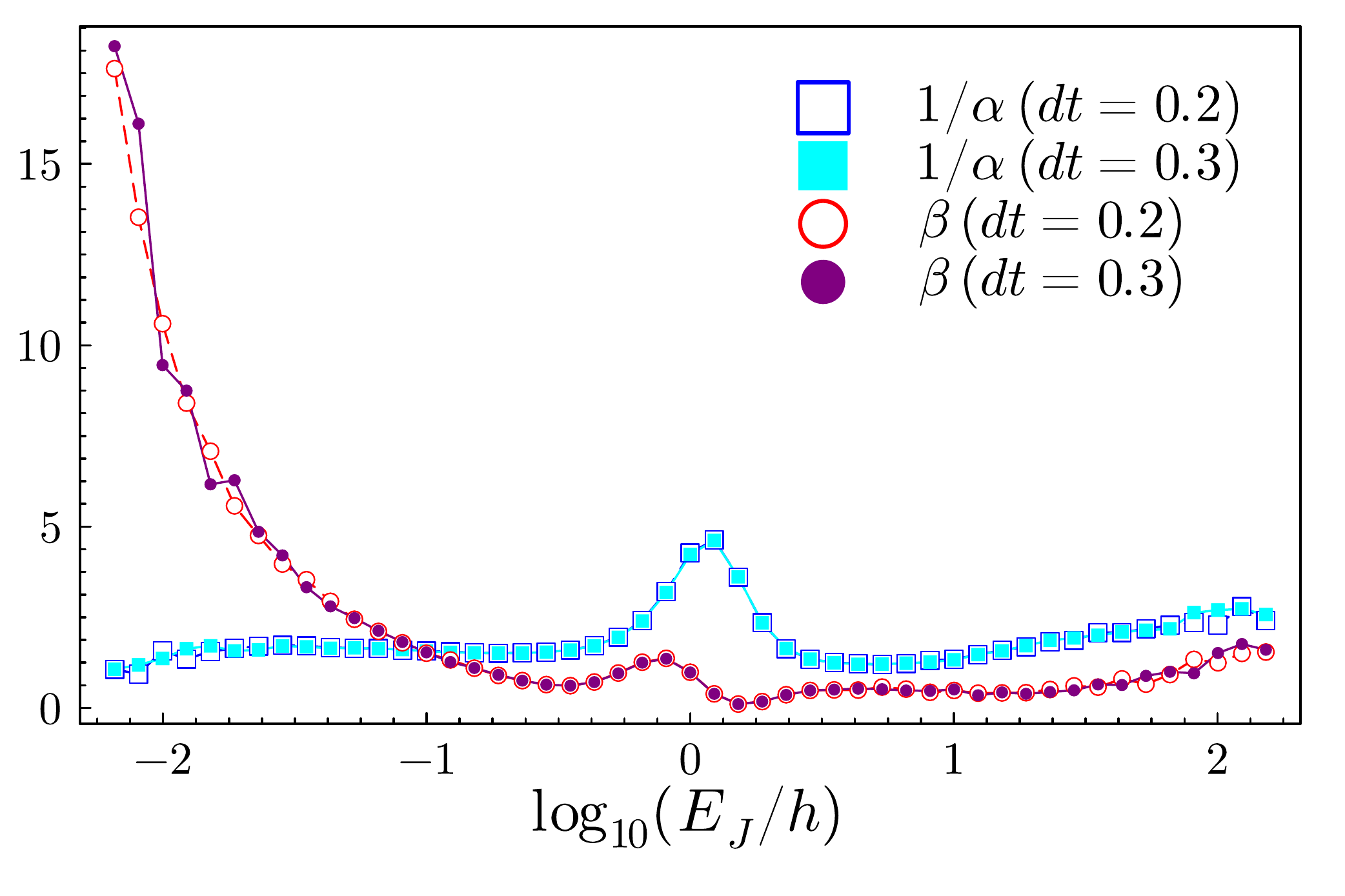}
    \caption{Fit coefficients for the ansatz in the main text for two different step sizes as a function of $E_J/h$.}
    \label{fig:as_and_bs_error}
\end{figure}

The Josephson-junction networks considered here are very likely non-stiff systems, although technically this could only be pinpointed by employing an adaptive-step method and analyzing step-size behavior for a given tolerance. The well-behaved drifts in Fig.~\ref{fig:errors} alone are not enough to determine numerical robustness when employing SIs \cite{benettin1994hamiltonian, hairer1997life}. It is, therefore, fundamental that results obtained with SIs be tested in order to exclude method- and step size-dependent effects. One option is to rerun a set of LS using a different (higher-order) SI, and another is fixing the algorithm and modifying the time step. Due to the extreme numerical sensitivity present in numerical simulations of strongly chaotic systems, this will generate a completely different trajectory even if the same initial conditions are used. Nevertheless, we must guarantee that the spectra at the end of the run are the same within the errors devised in the earlier section.  

We therefore retain the optimized 2nd method, but \emph{increase} the step size from $\text{d}t=0.2$ to $\text{d}t=0.3$. This is a particularly stringent test because we are essentially showing that the step size used in the simulations was more than enough to achieve robust results. A qualitative comparison between a set of mean LS in the SRN and LRN regimes obtained using different step sizes can be seen in Fig.~\ref{fig:convergence_spectra}. As is clearly visible, LS coincide within their error bounds for all values of $E_J/h$. For a quantitative comparison, we define the \textit{spectral error} for each $E_J/h$ value as the
root mean square
\begin{equation}\label{eq:se}
    \varepsilon_1(E_J/h) = \sqrt{ \frac{1}{N^2} \sum_{i=1}^{N^2} \left( \overline{\Lambda_i'} - \overline{\Lambda_i} \right)^2 }\quad,
\end{equation}
where the prime denotes a spectrum computed using a different step size. We then compare this error measure with the integrated \textit{combined error} for each curve, \textit{i.e.}
\begin{equation}\label{eq:ce}
     \varepsilon_2(E_J/h) = \sqrt{ \frac{1}{N^2} \sum_{i=1}^{N^2} \left[ \varepsilon(\overline{\Lambda_i'})^2 + \varepsilon(\overline{\Lambda_i})^2 \right] } \quad ,
\end{equation}
where $\varepsilon(\overline{\Lambda_i'})$ and $\varepsilon(\overline{\Lambda_i})$ are the point-errors in LS computed using $\text{d}t=0.3$ and $\text{d}t=0.2$. Stating that $\varepsilon_1 < \varepsilon_2$ means that changing the step size generates the same LS from the point of view of numerical accuracy, and in Fig.~\ref{fig:convergence_error} one can see this is precisely the case except for the strongly chaotic neighborhood of $E_J/h \approx 1$. However, the distance between the errors is smaller than $10^{-3}$, such that even here we can consider runs with different step sizes to have produced the same LS.  

As a last comparison, in Fig.~\ref{fig:as_and_bs_error} we display the fit coefficients for the ansatz used in the main text for the LS computed with different step sizes. It is clear that the power-law behavior for $\text{d}t=0.2$ is the same as for $\text{d}t=0.3$, and the critical exponents for both $\beta$ curves are the same within their error bounds. 

\bibliography{bib}